# Derivation of the Special Theory of Relativity from Invariance of Action

V. Hushwater[a)]

*70 St. Botolph St, apt. 401, Boston, MA 02116*

It is shown that kinematics of special relativity follows from postulating that action of a free particle's motion is invariant under the transformation from one inertial reference frame to another. We also assume that all such reference frames are equivalent and take into account that there exist massless particles (photons) that have changeable (not identically zero) momentum and the kinetic energy.

**I. INTRODUCTION**

Special theory of relativity (STR) is a well-established theory. Its kinematics and basic properties were derived by A. Einstein in 1905 from *the principle of relativity* (the principle that all inertial reference frames (IRFs) are equivalent and that the speed of light is the same in all such frames).[1]

However attempts to refine A. Einstein's derivation or even to replace the requirement of invariance of speed of light as the basic principle continue.[2] The motivation for this is psychological: as R. Feynman nicely explained, "Theories of the known which are described by different physical ideas may be equivalent in all their predictions, and are hence scientifically indistinguishable. However, they are not psychologically identical when trying to move from that base into the unknown. For different views suggest different kinds of modifications which might be made, and hence are not equivalent in the hypothesis one generates from them in one's attempt to understand what is not yet understood."[3,4]

Following that trend I show below that it is not necessary to consider the invariance of speed of light as the basic principle – it speaks only about a special, extreme type of motion. Instead, let us consider as the basic principle a relativistic property of a characteristic which is relevant to any motion – to action, which is an invariant in relativistic physics. Action is a major integral characteristic of mechanical processes – equations of the motion follow from the principle of the stationary action. In customary presentations of STR the invariance of action is just a consequence of the principle of relativity.[5] I will show however that the order can be reversed, that the invariance of speed of light can be viewed as a consequence of the principle that action is invariant.

As in classical mechanics and in the conventional approach to relativistic mechanics, I assume that all IRFs are equivalent. I consider light as a stream of photons – massless particles moving (in the geometrical optics approximation) along "light rays,"[5, 6] and I take into account that photons participate in mechanical processes and therefore have changeable (not identically zero) momentum and the kinetic energy.

I make my discourse within the framework of Lagrangian mechanics, which follows from the principle of stationary action. However, since this topic is not widely known I discuss in short its basics in the next section.

## II. LAGRANGIAN MECHANICS

Let us review, for a further use, foundations of mechanics in the Lagrange's formalism.[5,7,8] We will consider for simplicity a particle moving in one dimension. By the definition action $S$ is,

$$S(t_1, t_2) = \int_{t_1}^{t_2} dt L(x, v, t), \tag{1}$$

Where $x$ and $v$ are correspondingly coordinate and velocity of a particle at the moment of time $t$. Here $L = L(x, v, t)$ is so-called Lagrange function.

As follows from the (Hamilton's) principle of stationary action, $\delta S = \delta \int_{t_1}^{t_2} dt L(x, v, t) = 0$ $L = L(x, v, t)$ satisfies the Euler-Lagrange equation of the motion,

$$\frac{d}{dt}\frac{\partial L}{\partial v} - \frac{\partial L}{\partial x} = 0. \tag{2}$$

So $L$ is a differentiable function of $x, v,$ and $t$.

Let us first consider a free particle in an IRF. Due to homogeneity and isotropy of space and time the Lagrange function for it can depend only on $v^2$, $L = L(v^2)$. In such a case, as follows from Eq. (2), $\frac{\partial L}{\partial v} = const$ and since it depends only on $v$ it follows that $v = const$ as well, i. e. free particle moves in an IRF with a constant velocity. So we have from Eq. (1),

$$S(v, \Delta t) = L(v^2) \Delta t, \tag{3}$$

where $\Delta t = t_2 - t_1$.

The momentum, $p$ of a particle is a quantity conserved due to homogeneity of space. In terms of $L$ its expression is,[8]

$$p = \frac{\partial L}{\partial v}. \tag{4}$$

The energy of a particle, $E$ is a quantity conserved due to homogeneity of time,[8]

$$E = pv - L, \tag{5}$$

In the case of a particle interacting with the scalar field its Lagrange function has the form,[5,8]

$$L = L(x, v, t) = L(v^2) - U(x, t), \tag{6}$$

where $U(x, t)$ is the potential energy of the particle in the field.

For such an $L$ the Euler-Lagrange equation (2) turns into,

$$\frac{dp(v)}{dt} = F, \tag{7}$$

where $F \equiv -\frac{\partial U(x,t)}{\partial x}$ is a force.

As follows from observations acceleration $\frac{dv}{dt}$ is always points in the direction of $F$. For a one-dimensional motion along $x$ axis this means that $\frac{dv}{dt}$ and F have the same sign. Taking into account that $\frac{dp(v)}{dt} = \frac{dp}{dv}\frac{dv}{dt}$ it follows from (7) that $\frac{dp}{dv} = F/\frac{dv}{dt}$ is always positive,

$$\frac{dp}{dv} > 0. \tag{8}$$

Eq. (8) shows that momentum of a particle is continuous (even differentiable) function of $v$ and it increases with increasing of velocity.

We have to expect also that,

$$p(0) = 0 \tag{9}$$

since in the case $v = 0$ there is no a privilege direction along which must point $p$.

In (nonrelativistic) classical mechanics Lagrange function, $L_{cl}$ of a free particle is,[8]

$$L_{cl} = \frac{1}{2}mv^2, \tag{10}$$

where constant $m$ is called mass. So $S_{cl}$ on a time interval $\Delta t$ is,

$$S_{cl}(v, \Delta t) = \frac{1}{2}mv^2 \Delta t. \tag{11}$$

Substituting $L$ from Eq. (10) into Eq. (4) we obtain,

$$p_{cl} = mv, \tag{12}$$

and Eq. (7) turns into the second Newton law, $m(\partial v/\partial t) = F$. Also, as follows from Eq. (12), $dp_{cl}/dv = m$. So we have from Eq. (8) that $m > 0$. Finally, as follows from Eq.10 and Eq. (5),

$$E_{cl} = \frac{1}{2}mv^2 + U(x,t), \tag{13}$$

## III. RELATIVISTIC KINEMATICS AS A CONSEQUENCE OF THE INVARIANCE OF ACTION

Using the formalism of Lagrangian mechanics I show below that the requirement that *action must be an invariant* in respect to (not specified) transformation from one IRF to another leads to a consequence that the speed of all massless particles – is one and the same in all IRFs.

The classical mechanical action (11) for a free particle is not invariant under the (Galilean) transformation from one IRF to another since velocity $v$ changes (by algebraic subtraction of the new IRF's velocity) while $\Delta t$ is considered the same in all IRFs. Also it is zero in a reference frame, where the particle is at rest, $v = 0$ while the invariance of action requires that in such a reference frame it must be nonzero unless it is zero in all reference frames.

For action to be invariant $\Delta t$ must be frame-dependent and $L_{cl} = (1/2)mv^2$ must be replaced by a certain Lagrange function $L(v^2)$ such that $L(0) \neq 0$,

$$S = L(v^2)\Delta t_v, \tag{14}$$

where $\Delta t_v$ is a time interval in a IRF, relative to which the particle moves with velocity $v$.

Due to equivalence of all IRFs $L(v^2)$ is the universal function. *We do not know from the beginning how $L(v^2)$ and $\Delta t_v$ must depend on v but their product must be independent on it. That will provide frame-invariance of S.*

Our task is determining $L(v^2)$. In order to do that let us consider some properties of it and of its derivative, $\partial L/\partial v$. Let us first compare the duration of the motion of a particle with nonzero mass in two IRFs – the first, in which the particle moves with the speed $v$ and the second, in which the particle is at rest. From the invariance of action, Eq. (14) we have, $L(v^2)\Delta t_v = L(0)\Delta t_0$. So

$$\Delta t_0 = \left(\frac{L(v^2)}{L(0)}\right)\Delta t_v. \tag{15}$$

Transformation rule of a coordinate *x* and time *t* in one IFR to *x'* and *t'* in another IFR, which moves with velocity v in respect to the first is a property of space-time and (similarly to Galilean transformation) must not depend on any mass. Because of that, the factor $L(v^2)/L(0)$ in Eq. (15) must depend only on $v^2$ but not on *m*. At the same time $L(0)$ is, due to dimensional considerations, a constant proportional to mass, *m*. That is why in order for $L(v^2)/L(0)$ be mass independent $L(v^2)$ must be proportional to *m* as well. More precisely, it must have a form,

$$L(v^2) \equiv mf(v^2), \tag{16}$$

where $f(v^2)$ is a certain function that has dimension of $[v^2]$ and is *m*-independent, such that $f(0) \neq 0$.

Such a $L(v^2)$ leads to a consequence that momentum (4) and the energy (5) of a free particle are proportional to the particle's mass. That is necessary for satisfying the correspondence principle, the requirement that for small speeds relativistic formulas for physical quantities reduce to corresponding formulas of classical mechanics.

Classical mechanics, as follows from formulas (12) and (13) does not permit for massless particles to have nonzero momentum and kinetic energy. However such particles – photons – exist, having changeable (not identically zero) momentum and the kinetic energy, and therefore take part in mechanical processes. Mechanics with invariant action admits such a possibility. From Eqs. (4)

and (16) it follows that momentum,

$$p(v) = m\frac{\partial f}{\partial v}. \qquad (17)$$

As we noticed above, after Eq. (8) momentum of a particle (with nonzero mass) must increase with increasing of velocity and be continuous function of it. For that, as follows from Eq. (17) $\partial f/\partial v$ must be a continuous increasing function of $v$.

Substituting $m = 0$ into Eq. (17) one can see that a massless particle can have a nonzero momentum, $p(v)$ but only at values of $v$, for which $\partial f/\partial v = \infty$.

From the mathematical point of view a continuous monotonically increasing function ($\partial f/\partial v$), has only one maximum (or infinite limit) at the biggest possible value of $v$, $v_{max}$, which may be finite or infinite. (The later would be the case e. g. for $f(v) \propto v^2$.) But $v_{max} = \infty$ would mean that in order to have nonzero momentum (and energy) a massless particle must move with the infinite velocity. This is unacceptable since *physical quantities cannot be equal to the (actual) infinity – they must be, in principle, measurable*. So $v_{max}$ must be finite. We will use below the customary notation, $v_{max} \equiv c$. According to Eqs. (9) and (17) $(\partial f/\partial v)|_{v=0} = 0$. So $\partial f/\partial v$ must be a monotonically increasing function of $v$, from 0 at $v = 0$ to $\infty$ at $v = c$.

But for a particle with nonzero mass $(\partial f/\partial v)|_{v=c} = \infty$ leads, using Eqs. (17) and (5), to the conclusion that its momentum and the energy are infinite at $v = c$. This means that a particle with nonzero mass can have speed in the interval $0 \le v < c$ but cannot reach $c$ since its momentum and the energy grow infinitely when $v \to c$.

In classical mechanics the Galilean transformation from one IRF to another, which moves relative to the first one, changes the particle's velocity, momentum, and energy.[8] In case of mechanics with invariant action we should expect similar changes occur for particles with nonzero mass, although the transformation law has to be different since particles' speeds are only in the interval $0 \le v < c$. For a massless particle, its momentum and energy should change as well. However, its velocity must not change since a massless particle can have nonzero momentum and energy only if it moves with speed *c. Transformation law from one IRF to another must provide invariance of velocity, which is equal to ±c. That conclusion is the basic principle of the special theory of relativity.*

The derivation of this invariance essentially completes our task. The rest of relativistic kinematics, including the so-called Lorentz transformation,[9] follows from the fact that speed of light (of photons and of all massless particles) is the same in every inertial frame. Let us briefly discuss few additional points and find function $L(v^2)$.

As we found above, $\partial f/\partial v > 0$ for $0 < v \le c$, therefore $\partial L/\partial v > 0$ as well and $L(v^2)$ increases on this speed interval. As follows from eq. (5), for $v \to 0$ $E(v) \to -L(0)$. Requirement that the energy must be positive leads to the condition, $L(0) < 0$. This in turn requires that $L(v^2) < 0$ as well, in order for $\Delta t_v$ and $\Delta t_0$ have the same sign (which preserves an order of events in different reference frames). So $L(0) < L(v^2) \leq 0$ and as follows from eq. (15), $\Delta t_0 < \Delta t_v$, which expresses the 'time dilation' phenomenon in a moving reference frame.

Let us now compare the description of the motion of a massless particle in two IFRs, $K$ using variables $x$ and $t$, and $K'$ using variables $x'$ and $t'$. The particle has the same speed $c$ in both frames. That is why if in time interval $\Delta t$ it covers distance $\Delta x$ in $K$ frame we have, $c^2 \Delta t^2 - \Delta x^2 = 0$ and, similarly, $c^2 \Delta t'^2 - \Delta x'^2 = 0$. The quantity

$$\Delta s \equiv \sqrt{(c^2 \Delta t^2 - \Delta x^2)} \tag{18}$$

is called the interval.[5] We see that if $\Delta s = 0$, $\Delta s' = 0$ as well, and therefore in such a case

$$\Delta s = \Delta s' \quad \text{or} \quad c^2 \Delta t^2 - \Delta x^2 = c^2 \Delta t'^2 - \Delta x'^2. \tag{19}$$

For the motion of a particle with nonzero mass $\Delta x = v \Delta t$ and since $v < c$

$\Delta s = \sqrt{(c^2 \Delta t^2 - \Delta x^2)} = \sqrt{(c^2 - v^2)} \Delta t \neq 0$. Similarly $\Delta s' \neq 0$ since $v' < c$. Nonetheless the equality (19) is satisfied in such a general case as well.[10] I. e. *the interval is invariant*.

If a particle moves with velocity $v$ relative to reference frame $K$ and is at rest in reference frame $K'$, $\Delta x' = 0$, $\Delta t' \equiv \Delta t_0$, $\Delta t \equiv \Delta t_v$ and $\Delta x = v \Delta t_v$. So we have from eq. (19),[5] $(c^2 - v^2)\Delta t_v^2 = c^2 \Delta t_0^2$ and

$$\Delta t_0 = \Delta t_v \sqrt{(1 - v^2/c^2)}. \tag{20}$$

Substituting formula (20) in Eq. (15) we find,

$$L(v^2) = L(0)\sqrt{(1 - v^2/c^2)}. \tag{21}$$

Let us check if this $L(v^2)$ has properties discussed above. We have, $\partial L/\partial v \propto v/\sqrt{(1 - v^2/c^2)}$. Thus $(\partial L/\partial v)|_{v=0} = 0$ and $(\partial L/\partial v)|_{v=c} = \infty$, as it must be.

In order to figure out $L(0)$ let us consider expansion of the Lagrange function, $L(v^2)$ for small velocity, $v \ll c$,

$$L(v^2) \cong L(0)(1 - v^2/2c^2 + \cdots). \tag{22}$$

Taking into account that in such a case formulas of special relativity for momentum (4) and the energy (5) must reduce to corresponding formulas of classical mechanics we must have

$$\text{if } v \to 0, \quad L(v^2) \approx L(0) + L_{cl}(v^2). \tag{23}$$

Comparing formulas (22) and (23) we obtain, $L(0) = -mc^2$. So

$$L(v^2) = -mc^2\sqrt{(1 - v^2/c^2)}. \tag{24}$$

Using this formula one can easily find from Eqs. (4) and (5) expressions for momentum and the energy in relativistic mechanics.[5] Also, substituting formula (24) in Eq. (3) we have,

$$S(v, \Delta t) = L(v^2)\Delta t = -mc\Delta s, \qquad (25)$$

as it must be[5] – two invariant quantities, the action, $S(v, \Delta t)$ and the corresponding interval, $\Delta s$ are proportional to each other.

## III. CONCLUSION

Thus we derived kinematics and formulas for basic physical quantities of the special relativity from the principle of the invariance of action within the framework of Lagrangian mechanics, taking into account few requirements like the correspondence principle etc.

J. Ziman wrote, [11] "Special theory of relativity can be logically substantiated by many ways that satisfy almost all tastes and worldviews." The worldview advanced in this paper is that *action is a major physical concept.* The most important physical concepts are those which have the same values in all IRFs, i.e. which are invariant under the transformation from one IRF to another. Such invariants give frame-independent, objective characteristics of a system.[12]

In connection with this it is interesting to notice that while local characteristics of a particle's motion – its space-time coordinates, velocity, momentum and energy etc. are frame-dependent, its integral (nonlocal) characteristics – action and interval are its frame-independent, objective characteristics.

## ACKNOWLEDGMENTS

I thank Peter Milonni for encouraging comments on the paper. I also acknowledge Ted Jacobson for his comments on the first draft of the paper.

[a)] Electronic mail: quantcosmos@gmail.com

explained all then known observations and experiments, see e. g. O. Darrigol, *Electrodynamics from Ampere to Einstein* (Oxford, N. Y., 2000); O. Darrigol, "The Genesis of the Theory of Relativity," in T. Damour et al eds., *Einstein, 1905-2005 (Poincaré Seminar 2005), Progress in Mathematical Physics,* **47**, *(BirkhäuserVerlag,* Basel, 2005)*,* 1-31.

5. L. Landau and E. Lifshitz, *The Classical Theory of Fields* (Pergamon Press, Oxford, and Addison-Wesley, Reading, MA, 1987).

6. Not only such "classical photons" but all particles in classical and relativistic mechanics are an approximate description of corresponding quanta. See e. g. E. Wichmann, *Quantum Physics* (McGraw-hill, N. Y., 1971). A. Einstein in Kinematical Part of his first paper, Ref. 1, in which he derived the Lorentz transformation, speaks only about light rays, there is no a word on light as electromagnetic wave.

7. D. Morin, *Introduction to classical mechanics, with problems and solutions* (Cambridge UP, Cambridge, 2008).

8. L. Landau and E. Lifshitz, *Mechanics* (Pergamon Press, Oxford, 1960).

9. "Lorentz transformation" of a coordinate $x$ and time $t$ from one IFR, $K$ to another, $K'$ ($x'$ and $t'$), which moves with velocity $v$ in respect to $K$ has the form: $x' = (x - vt)\gamma$, $t' = (t - v/c^2 x)\gamma$, where $\gamma = 1/\sqrt{(1 - v^2/c^2)}$. Its exact form was published before the theory of relativity by J. Larmor in 1897 and by H. Lorentz in 1899 as the transformation, which leaves the wave equations of the free electromagnetic field form-invariant, see e.g. in M. Macrossan, "A note on relativity before Einstein," Brit. J. Phil Sci. **37**, 232-234 (1986). Lorentz transformation was rederived on the basis of the principle that speed of light is invariant by A. Einstein in 1905, see Ref. 1.

10. Invariance of the interval in general case can be easily proved by e. g. by using the Lorentz transformation from Ref.9.

11. J. Ziman, *Elements of Advanced Quantum Theory* (Cambridge, Cambridge UP, 1969) Ch. 6, Sec. 1.

12. M. Born, "Some Philosophical Aspects of Modern Physics," Proc. Roy. Soc. Edin, **57**, 1-18 (1936), reprinted in *Physics in my generation*, 2nd edit. (Springer, London, 1969); M Born, "The concept of reality in physics," *Bulletin of the Atomic Scientists*, **14**, 313-321 (1958).